\documentstyle[12pt,epsfig]{article}
\def\ifmath#1{\relax\ifmmode#1\else$#1$\fi}

\def\ie{{\it i.e.}}

\def\ra{\rightarrow}
\def\bc{\begin{center}}
\def\ec{\end{center}}
\newcommand{\beq}{\begin{equation}}
\newcommand{\eeq}{\end{equation}} 
\newcommand{\beqa}{\begin{eqnarray}}
\newcommand{\eeqa}{\end{eqnarray}}

\def\CP     {\ifmath{C\!P}}

\def\lbabar{\mbox{\sl {\Large\sl B}\hspace{-0.45em} A\hspace{-0.1em}{\Large\sl B}\hspace{-0.45em} A\hspace{-0.1em}R}}
\def\babar{\mbox{\sl B\hspace{-0.4em} {\scriptsize\sl A}\hspace{-0.4em} B\hspace{-0.4em} {\scriptsize\sl A\hspace{-0.1em}R}}}
\newcommand{\babarnote}[1] {\mbox{\babar\ {\sl Note}~\# #1}}

\newcommand{\np}   {{\sl Nucl. Phys.~}}

\newcommand{\prl}  {{\sl Phys. Rev. Lett.~}}
\newcommand{\pr}   {{\sl Phys. Rev.~}}

\def\epem  {\ifmath{e^+e^-}}

\def\KS    {\ifmath{K^0_{\scriptscriptstyle S}}} 
\def\KL    {\ifmath{K^0_{\scriptscriptstyle L}}} 

\def\bz    {\ifmath{B^0}}
\def\bzb   {\ifmath{\overline{B}{}^0}} 

\def\BzBzb {\ifmath{B^0 \overline{B}{}^0}} 
\def\Kbar{\ifmath{\kern 0.2em\overline{\kern -0.2em K}}{}}   
\def\Bbar{\ifmath{\kern 0.18em\overline{\kern -0.18em B}}{}} 
\def\Dbar{\ifmath{\kern 0.2em\overline{\kern -0.2em D}}{}}   
\def\jpsi  {\ifmath{{J\mskip -3mu/\mskip -2mu\psi\mskip 2mu}}}
\def\FourS {\ifmath{\Upsilon{\rm( 4S)}}} 
\def\Y#1S{\ifmath{\Upsilon\rm(#1S)}} 

\def\UT{Unitarity Triangle}

\def\kev  {\ifmath{\mbox{\,ke\kern -0.08em V}}} 
\def\mev  {\ifmath{\mbox{\,Me\kern -0.08em V}}} 
\def\gev  {\ifmath{\mbox{\,Ge\kern -0.08em V}}} 
\def\gevc {\ifmath{\mbox{\,Ge\kern -0.08em V$\!/c$}}} 
\def\mevc {\ifmath{\mbox{\,Me\kern -0.08em V$\!/c$}}} 
\def\gevcc{\ifmath{\mbox{\,Ge\kern -0.08em V$\!/c^2$}}} 
\def\mevcc{\ifmath{\mbox{\,Me\kern -0.08em V$\!/c^2$}}} 

\def\mum  {\ifmath{\,\mu\mbox{m}}}

\def\CsI{CsI(Tl) }

\newif\ifdimspec

\def\checkdim#1{\ifx#1\end \let\next=\relax
  \else \ifcat#1a \dimspectrue \fi \let\next=\checkdim\fi \next}
\newcommand{\fallblcaption}[2]{\caption{\sl #2\label{fig:#1}}}
\newcommand{\FigEPS}[3]
{
\begin{figure}[htbp] 
\begin{center}
  \mbox{\epsfig{file=#1.eps,width=#3}}
\end{center}
\fallblcaption{#1}{#2}
\end{figure}
}

\newcommand{\FigPS}[3]
{
\begin{figure}[htbp] 
\begin{center}
\mbox{\epsfig{file=#1.ps,width=#3,bbllx=23pt,bblly=180pt,bburx=516pt,bbury=683pt}}
\end{center}
\fallblcaption{#1}{#2}
\end{figure}
}

\newcommand {\dmd}{{\Delta {m}_{d}}}
\newcommand {\dms}{{\Delta {m}_{s}}}

\newcommand {\vbu}{{V}_{ub}}
\newcommand {\vbc}{|V_{cb}|}
\newcommand {\vbuvbc}{|\frac{\vbu}{V_{cb}}|}

\newcommand {\epsk}{|\varepsilon_K|}
\newcommand {\fbb}{f_{B_d} \sqrt{B_{B_d}}}
\newcommand {\fbs}{f_{B_s} \sqrt{B_{B_s}}}

\newcommand {\ssa}{\sin 2 \alpha}
\newcommand {\ssb}{\sin 2 \beta}

\def\bzbzbar{{\hbox{$B_0$}}{\hbox{\kern -.5cm\lower -.45cm \hbox{${\scriptscriptstyle(}{\scriptstyle -}{\scriptscriptstyle )}$}}}}
\def\rhorhobar{{\hbox{$\rho$}}{\hbox{\kern -.65cm\lower -.45cm \hbox{${\scriptscriptstyle(}{\scriptstyle -}{\scriptscriptstyle )}$}}}}

\def\lf {{ \lambda_f}}

\def\cmd {{\cos (\dmd t)}}
\def\smd {{\sin (\dmd t)}}
\def\kin {\ifmath{\dot{\kappa}}}
\def\KIN {\ifmath{\cal KIN}}

\begin{document}
\begin{titlepage}
\begin{flushright}
{\bf
LAL 98-16}\\
\end{flushright}
\vskip 2cm 
\begin{center}
{\large\bf \CP~violation with \lbabar} \\

\vspace{2.cm}
{\large St\'ephane Plaszczynski \footnote{e-mail:plaszczy@lal.in2p3.fr}} \\
\vspace{.5cm}
{\sl Laboratoire de l'Acc\'el\'erateur Lin\'eaire,\\
     IN2P3-CNRS et Universit\'e de Paris-Sud, F-91405 Orsay}
\vfil
\begin{abstract}

The \babar~experiment is a new generation detector located at the SLAC
B factory PEP-II ring which should start taking data at the end of
1999. Its main goal is the study of \CP~violation in the \BzBzb~
system. After explaining the nature of this \CP~violation, I review 
the scientific program for achieving this study in many different
modes, in the light of the recent developments obtained both on the
experimental and theoretical side. Implications for the Standard
Model are then discussed.

\end{abstract}
\vspace{1.5cm}
{\it Invited talk at the $XII^{th}$ Rencontres de physique de la Vall\'ee
d'Aoste, la Thuile, Italy, March 1-7 1998 }

\end{center}
\end{titlepage}

\section{Introduction}

So far the violation of \CP~symmetry has just been observed in the
neutral Kaon sector. The Standard Model can accommodate for such a
violation, through the CKM mixing matrix. Furthermore, it even {\it
predicts} \CP violation in the \BzBzb~ system. A first task of a B
factory is thus to check whether such a prediction holds on in the \bz~
sector. 

The CKM matrix is presently one of the less tested
sector of the Standard Model. Indeed, two of its four parameters 
are presently very badly known ( $\rho$ and $\eta$ in the Wolfenstein 
parametrization).
The knowledge on these two parameters is depicted in
the so-called Unitarity Triangle (UT), where the apex of the scaled triangle 
is precisely the $\rho,\eta$ point (Fig \ref{fig:triangle}).
\FigEPS{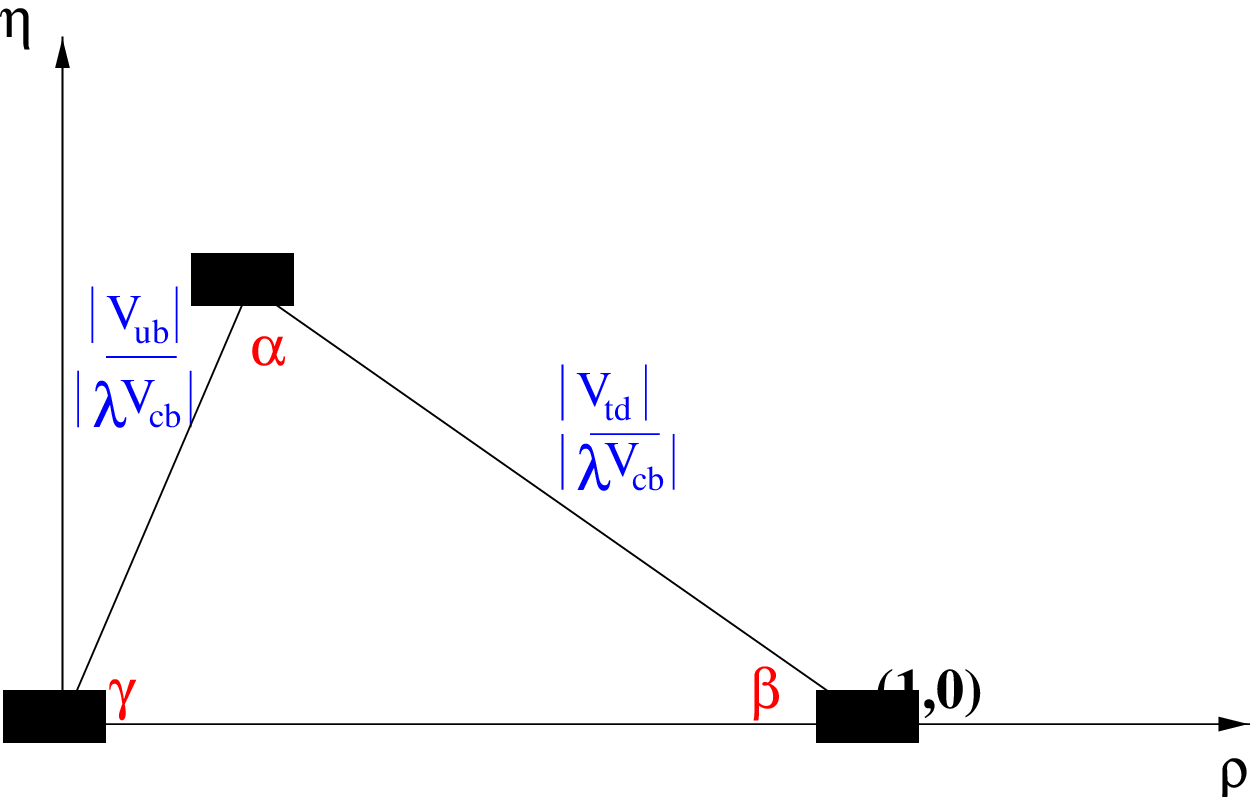}{The Unitarity Triangle representation. Note that its
base is normalized to 1}{3.5in}

Presently the three {\it sides} of the triangle are measured, but the
extraction from the data of CKM quantities requires the knowledge of
model-dependent theoretical parameters (coming from non-perturbative
QCD and models for heavy to light transitions). Another
constraint comes from the measurement of \CP~violation in the kaon
system, but here again, due to large theoretical uncertainties, this
constraint is quite weak. The use of limits on the $B_s$ mixing
frequency is interesting but again plagued by a theoretical parameter 
(section \ref{sec:sm}).

The goal of B factories is to measure two {\it angles} of the UT ($\alpha$ and
$\beta$) in a clean way. 
Combining all observables will allow to (over-?)constrain the CKM
matrix.
Furthermore, time dependent \CP~violating asymmetries, being rare
processes, are sensitive to New Physics phenomenon. Or, said in
another way, many extensions
of the Standard Model includes some new sources of \CP~violation 
\cite{nirNP} that could be observed at a B-factory experiment.

\section{Which \CP~violation?}
In the $\FourS \ra \BzBzb$~ decay, after the decay of a tagging flavor B,
the time distribution of the decay of the other B (to a final state $f$) 
is of the form:
\beqa
\label{eq:rho}
\rho(t)&=&C e^{-\Gamma t} [ 1+|\lf|^2 {+}
(1-|\lf|^2)\cmd {-} 2 Im \lf \smd] ~\mbox{(\bz tag)} \nonumber \\
\bar{\rho}(t)&=&C e^{-\Gamma t} [ 1+|\lf|^2 {-}
(1-|\lf|^2)\cmd {+} 2 Im \lf \smd] ~\mbox{(\bzb tag)}
\eeqa
where
$\lf = \frac{q}{p}\frac{{\cal A}(\bzb\ra f)}{{\cal A}(\bz\ra f)} $
\footnote{$q,p$ appear in the physical states decomposition:
$|B_L\rangle= p |\bz\rangle + q |\bzb\rangle$ and $|B_H\rangle= p
|\bz\rangle - q |\bzb\rangle$.},
$\dmd$ is the $B_d$ mixing frequency and $\Gamma$ its width.
 One notices the difference in signs in
the above expression.

Choosing a final \CP~ eigenstate, the time dependent asymmetry $a(t)$ can be
different from 0, indicating \CP~ violation:
\beq
\label{eq:a}
a(t) = \frac{N(\bz(t)\ra f)-N(\bzb(t) \ra {f})}{N(\bz(t)\ra f)+N(\bzb(t) \ra
{f})} = \frac{ (1-|\lf|^2)\cmd -2 Im \lf \smd}{(1+|\lf|^2)}
\eeq

There are two ways for the ratio to be non zero:
\begin{itemize}
\item $|\lf| \neq 1$ \\
This can be achieved either by $|\frac{q}{p}| \neq 1$ or $\frac{{\cal
A}(\bzb\ra f_{CP})}{{\cal A}(\bz\ra f_{CP})} \neq 1$. The former
inequality represents a \CP~violation in the mixing (indirect) and the
latter a \CP~violation in the decay (direct). The amount of indirect
\CP~violation is expected to be very
small in the \BzBzb~system (at a level of $10^{-3}$). Direct
\CP~violation  however can be different from 0 in rare processes
(beyond tree diagrams) and depends on the modes studied. 
\item $Im \lf \neq 0$ \\
The term $Im \lf$ has no reason to be equal to 0. In some ``clean''
cases it can even be directly related to the angles of the \UT: $Im
\lf =\ssa$ or $Im\lf =\ssb$. It arises from the interference between
the decay with and without mixing. It is the prime motivation for the
construction of B-factories.
\end{itemize}

\section{\label{sec:req} Introducing \babar}
To achieve an experimental study of such time dependent asymmetries,
the following requirements must be fulfilled:
\begin{itemize}
\item produce a coherent \BzBzb state (\ie~run a the \FourS resonance).
\item since CP modes are rare (BR of the order of $10^{-4},10^{-5}$)
have a high luminosity,
\item the time variable that appears in Eq.(\ref{eq:a}) being the decay
time between the two B decays ($t=\Delta t$), 
it is crucial that they do not decay at
the same point (otherwise the $\smd$ term cannot be measured in a time
dependent way and the time-integral over this term vanishes): one needs
therefore to boost the \BzBzb~ system, \ie~use asymmetric beams.
\end{itemize}

The \babar~detector is located at the PEP-II storage ring, 
a high luminosity collider 
(${\cal L}= 3.10^{33} cm^{-2}s^{-1}$ is expected) of 9 \gev~electrons
against  3.1~\gev positrons. This gives a boost to the \BzBzb~system of
$\gamma \beta = 0.56$; the mean separation between the two B decays is
about $260 \mum$.

The (asymmetric) detector is a classical one for \epem~colliders 
(except for the DIRC), made of high quality components. Going from
the beam pipe (Fig \ref{fig:detector}):
\begin{itemize}
\item a 5 layer {\bf silicon vertex tracker} 
\item a low density He-based {\bf Drift chamber}
\item a \CsI {\bf calorimeter} with high granularity.
\item a {\bf DIRC} (Detection of Internally Reflected Cerenkov light)
for particle identification
\item a superconducting coil of 1.5 T.
\item An {\bf instrumented flux return} optimized for $\mu$ and \KL~detection.
\end{itemize}

\FigEPS{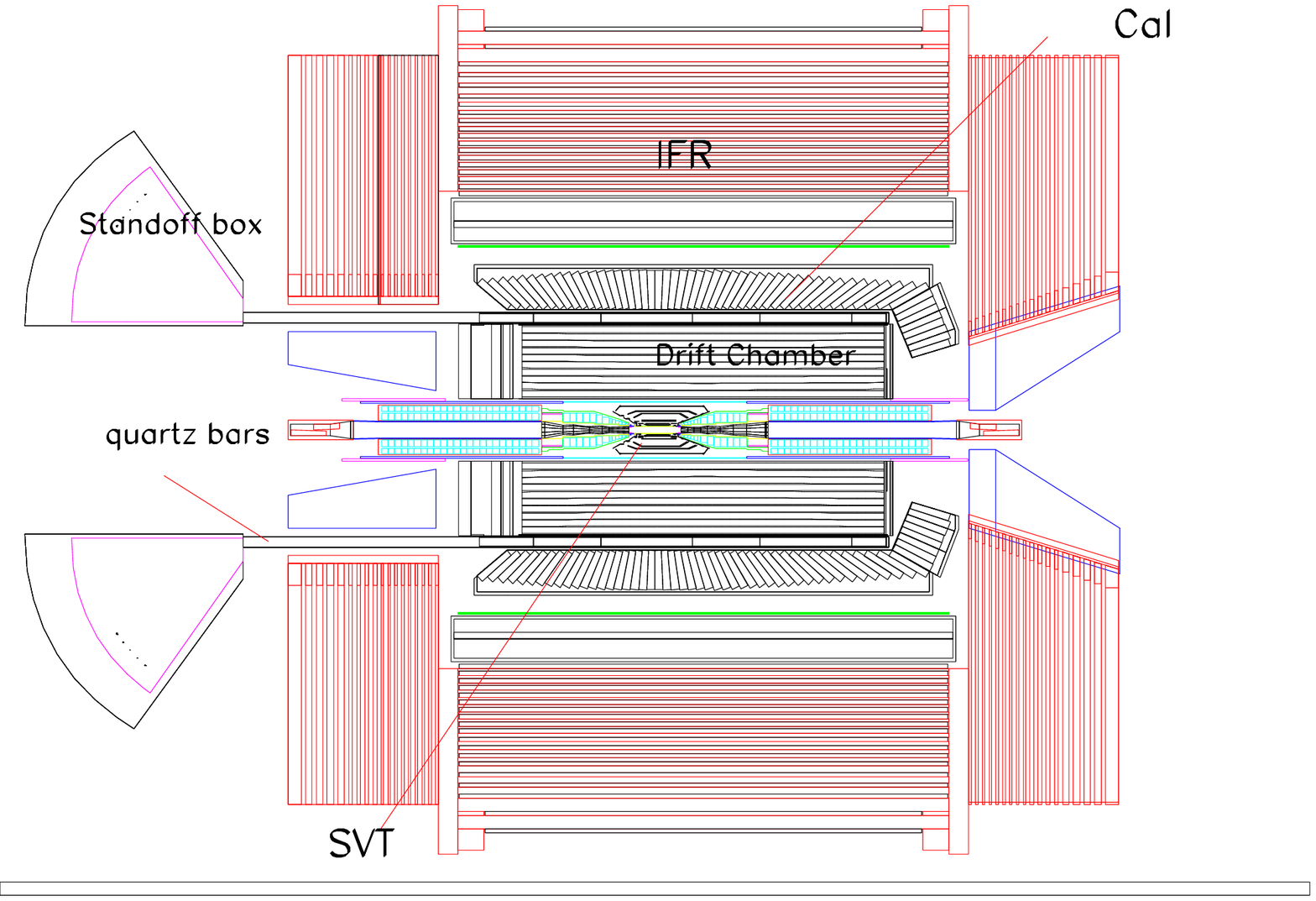}{Sketch of the \babar~detector}{7 in}
The {\bf DIRC} is a new detector for Particle Identification based on the
Cerenkov light emission of a particle passing through a quartz
bar. While generally the light captured in the radiator is lost, here
one uses this component which is trapped inside the quartz bar and
propagate by internal reflection to the end, conserving its
characteristic angle. At the end of the bar, the photon
propagates into a large volume of water (the ``standoff box'') and
reaches a huge array of about 13 000 photomultipliers. The
reconstruction of the angle between the hit PMT and the bar allows
to measure the Cerenkov angle, and thus the nature of the track.

\section{The \CP~program}

In order to extract \CP~ violation parameters, one needs in real life
to perform the following program:

\subsection{Reconstruction of a final state $f$}
This is performed by usual techniques (mass peaks...), for many
different final states:
\beq
f = \jpsi \KS,~\pi^+\pi^-,~\pi^+\pi^-\pi^0,~ 
4 \pi, D^+ D^-, \jpsi K^{\ast}, D^{\ast +}D^{\ast -}, \psi(2S) K,..
\eeq
I will detail the first three modes while describing the following
of the analysis.

\subsection{Tagging}
The goal of this part is to tag the flavor of the B meson ($b$ or
$\bar b$ quark?). This is generally performed searching for a lepton and/or
kaon in the event (Fig. \ref{fig:tagging})
\FigEPS{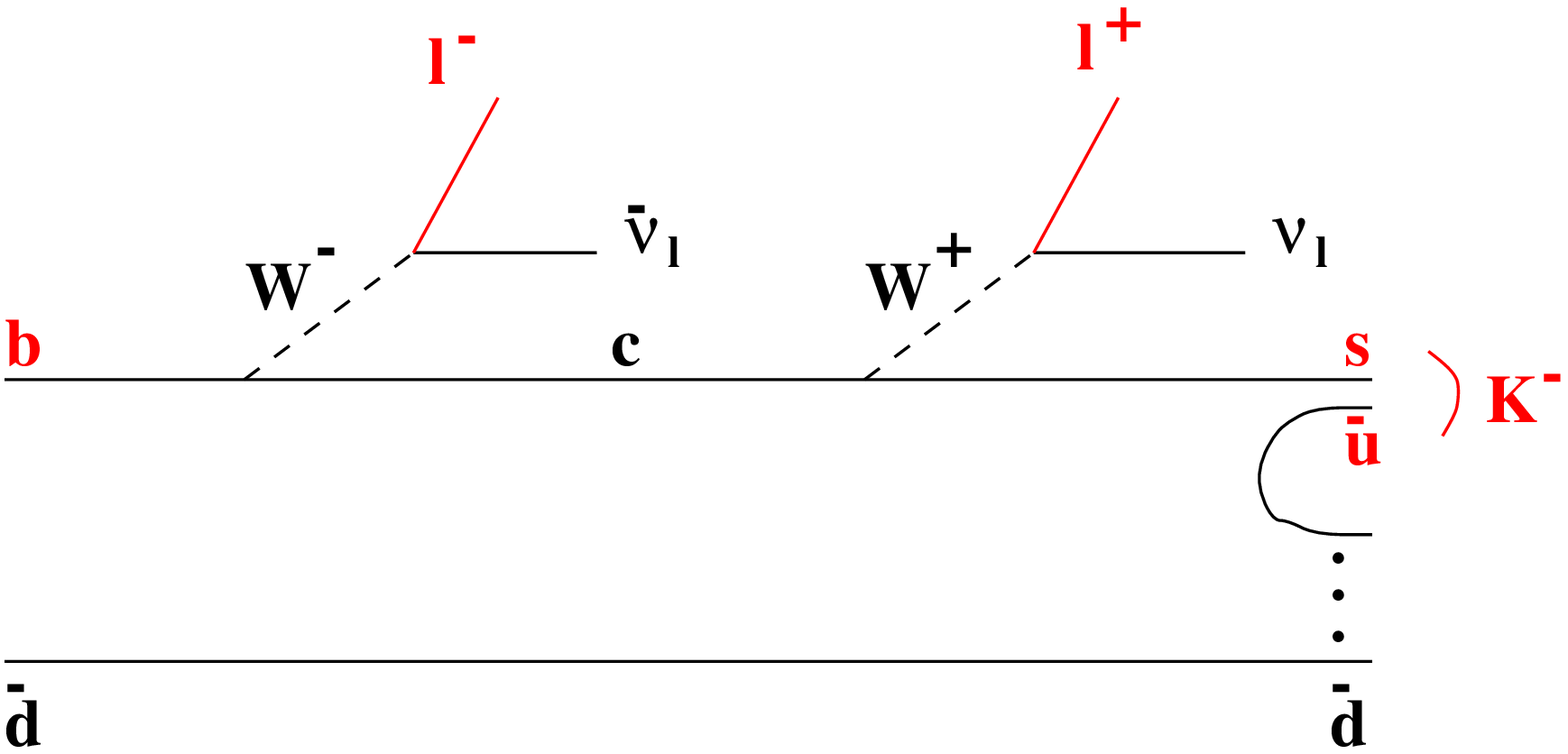}{Ways to produce a lepton and a kaon in a \bzb
decay}{4 in}
A sign contamination comes from secondary leptons; usually one uses a
cut (as on its momentum) to enrich the sample in primary
leptons. However in that case, one looses the information
contained in the secondary leptons: if it is very soft, it is more
{\it likely} to be a secondary lepton, so its sign information should
be flipped.

A tool named $CORNELIUS$ \cite{physbook} has been developed in the 
Collaboration, in order to {\it combine} the information of many
discriminating variables associated to the lepton. 
This is achieved using various multivariate
methods \footnote{presently it incorporates a Likelihood analysis, a
Fisher discriminant and a Neural Network}; it allows to crosscheck
the different outputs and have a grip on systematics.
But much more. It allows to assign to each event a {\it probability} to come
from a $b$ or $\bar b$ quark. This probability is then input in the
final likelihood determining the asymmetry and exploits optimally all
the available information.

The deterministic ``cut'' method degrades the determination on the
asymmetry by a dilution $D=\epsilon(1 -2 \mbox{w}^2)$ where $\epsilon$ is
the tagging fraction and w the mistag fraction. Previous estimates of
this quantity \cite{tdr} gave about: $D=0.33$. Using the probability
method allows to reach $D=0.41$, by combining 8 discriminating
variables for the lepton.

Notice that this combination can also be used to reject the 
background (generally ``continuum background'') by combining
discriminating variables based on the event topology. $CORNELIUS$ can
provide event by event a probability to be a $B \bar{B}$ or $q \bar q$
event.

\subsection{Time determination}
Once a mode is reconstructed, a vertex is performed with the remaining
charged tracks. The difference in space between both vertices
represents simply $\gamma \beta c t$ where $\gamma \beta$ is
the known boost of the machine (=0.56). The resolution on the
distance between both vertices is, for the \jpsi\KS~mode, about
50\mum, well beyond the mean $260\mum$ quoted in section
\ref{sec:req} for the mean B separation.

\subsection{Extracting \CP}
There are two aspects in extracting a quantity relevant for
physics. The first one is mainly experimental and is based on the
knowledge of the detector. It will be illustrated on the $\jpsi \KS$
mode. Going from a measured asymmetry to a relevant CKM quantity is a
more theoretical problem, that will be illustrated on the $\pi^+\pi^-$
mode.

\subsubsection{Experimental side: introducing the \KIN~variable
(\jpsi \KS)}

The extraction of the asymmetry can be performed by a likelihood fit to
the observed events. However it is more convenient to use the \KIN~variable.

In a simple case (the theoretically clean mode :$\jpsi \KS$) and
neglecting for the time being detector effects, the event distributions
(\ref{eq:rho})~can be written:
\beqa
\bz~tag:\rho(t)& = & C e^{-\Gamma t} [ 1 {-} \ssb \smd] \nonumber \\
\bzb~tag:\bar{\rho}(t)& = &C e^{-\Gamma t} [ 1 {+} \ssb \smd]
\eeqa

Constructing event by event the asymmetry
\beq
S_{tag} \frac{\rho -\bar{\rho}}{\rho +\bar{\rho}}
~=S_{tag}\smd\ssb= \kin \ssb 
\eeq
(where $S_{tag}=\pm1$ for \bzbzbar~~tag) 
allows to fill an histogram of the \KIN~($\kin$) variable (see  Fig
\ref{fig:kinideal}).

\FigEPS{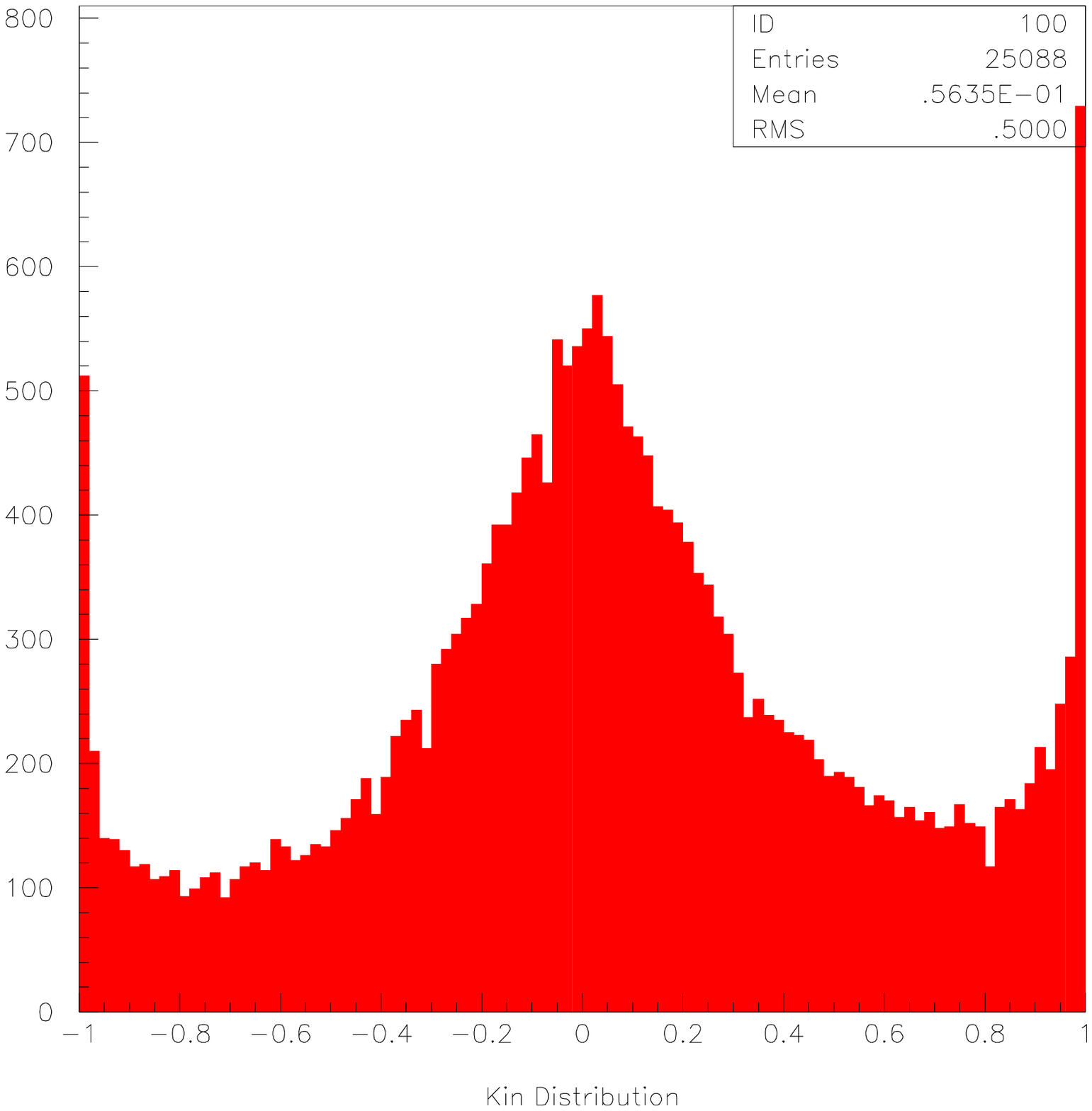}{An example of the \KIN~distribution on $\jpsi \KS$
mode. From the upper box, one can get the whole information on the
determination of $\ssb$}{3in}
\vskip -1cm

The distribution of $\kin$ variable has the nice property that, to a
very good approximation, one can get the estimate of the asymmetry
($\ssb$) by the very simple formula \cite{kin}:
\beq
\ssb \simeq \frac{\langle \kin \rangle}{\langle \kin^2 \rangle} 
\pm \frac{1}{\sqrt{ N \langle \kin^2 \rangle}}
\eeq

This means that a single plot (as Fig \ref{fig:kinideal}) carries the
whole asymmetry information, and that the asymmetry measurements can be
obtained via the number of entries, the mean and the RMS of
this histogram. Furthermore one can incorporate in the \KIN~, 
the tagging probability event by event and the time resolution
measurement \cite{kin}. All the \KIN~results still hold.
Finally notice, that the different modes can be combined in a
straightforward way by just summing the $\kin$ histograms.

Using the \KIN~approach , a recent analysis of the \jpsi \KS mode has
been performed \cite{remi}. The measurement obtained for $30
fb^{-1}$ (one ``nominal'' year) is:
$\ssb = 0.82 \pm 0.15$ (while .70 was generated).

\subsubsection{Theoretical side: the penguin world ($\pi^+\pi^-$)}

Contributing to a final state as $\pi^+\pi^-$, can exist, beside the
Tree Cabbibo suppressed mode (a), some modes as ``penguin
diagrams''(b). Recently emphasis has also been put on 
``long distance penguins''(c) (or ``charming'' penguins) which are of
QCD non-perturbative nature and results from
annihilation/re-scattering processes.

\FigEPS{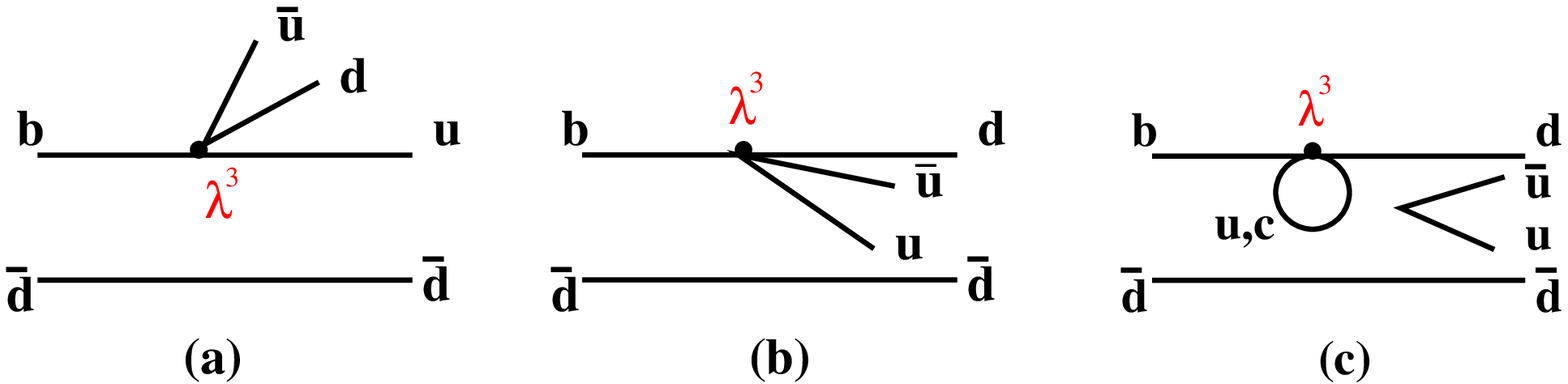}{Possible contributions to the $\pi^+\pi^-$ mode, in
term of local 4-fermion operators: (a) tree (b)penguin (c)
``charming'' penguin}{5in}

The theoretical estimate of the penguin contribution is very
delicate (and model-dependent). For this mode, it is expected that the
penguin contribution is ``smaller'' than the Tree one. However, recent
CLEO measurements of $BR(\bz\ra K \pi)$ \cite{BKPI} indicates that these penguin
modes indeed exist and should not be neglected in extracting the CKM
quantity $\ssa$.

Under the influence of penguins, the time distribution of the events 
Eq.(\ref{eq:rho}) can be written as:
\beq
 C e^{-\Gamma t}(1 \pm {R} \cmd \mp  \sin 2 ({\alpha + \delta}) \smd)
\eeq
where $R$ is the amount of direct \CP~violation, $\delta$ is a shift
due to the presence of penguins and $\alpha$ is the CKM angle.

What one can extract from the data is a $\sin 2 \alpha_{eff}$ but the
penguin shift is unknown. There are several solutions to this
problem, depending on what will be measured:

\begin{itemize}
\item Gronau and London \cite{gronau} have shown that measuring the
decoupled amplitudes: 
\beqa
A(\bz\ra\pi^+\pi^-)&,&A(\bzb\ra\pi^+\pi^-) \nonumber \\
A(B^+\ra\pi^+\pi^0)&,&A(B^-\ra\pi^-\pi^0)  \\
A(\bz\ra\pi^0\pi^0)&,&A(\bzb\ra\pi^0\pi^0) \nonumber 
\eeqa
allows to extract $\alpha$ by using the isospin symmetry. This is
however performed with an 8-fold ambiguity on $2 \alpha$. Furthermore,
the amplitudes of the $\pi^0\pi^0$ mode are expected to be small
(color suppressed) and experimentally difficult to reconstruct.
\item In the case where only an upper bound has been obtained for this
 $\pi^0\pi^0$ mode, Grossman and Quinn \cite{grossman} have shown
that the penguin shift is limited by:
\beq
{ \sin^2\delta} \le \frac{BR(\bz\ra\pi^0\pi^0)+BR(\bzb\ra\pi^0\pi^0)}
{BR(B^+\ra\pi^+\pi^0)+BR(B^-\ra\pi^-\pi^0)}
\eeq
This can be very useful when constraining the penguin effects.

\item Finally one will have to rely on the theorist understanding of
the penguin , reducing the model-dependence to a minimum number of
parameters \cite{charles}, to obtain a systematic error on the
determination of $\ssa$.
\end{itemize}

\subsubsection{The full problem: $\pi^+\pi^-\pi^0$}

A clearly challenging mode to extract $\alpha$ 
is $\pi^+\pi^-\pi^0$. In that case the situation is complicated by the
fact that:
\begin{itemize}
\item $\rho-\pi$ is assumed to dominate but $\rho^+ \pi^-, \rho^-
\pi^+,\rho^0 \pi^0 $ interferes.
\item Experimentally the signal is not so pure (signal:background
$\simeq$ 1:1)
\item There is a unknown contribution from penguins.
\end{itemize}

However this mode is important since, it is expected to have a higher
branching ratio than $\pi^+\pi^-$.
Also since it is a non-CP final state (due to phase space) it can have
a large $\cmd$ contribution, leading to a simultaneous determination of
$\cos 2\alpha$ and $\ssa$. This would definitely reduce the ambiguities due
to a single measurement of $\ssa$ (in which case $2 \alpha=\arcsin (\ssa)$ and 
$2 \alpha=\pi-\arcsin (\ssa)$ are both solutions).

The observed asymmetry in this case is of the form:
\beq
 C e^{-\Gamma t}[1 \pm {b(\Phi)} \cmd \mp  {c(\Phi)}  \smd]
\eeq

where $b(\Phi),c(\Phi)$ are {\it functions} of the phase space
(as the 2 Dalitz plot coordinates).

If one collects enough data, the study of the time-dependent Dalitz
plot allows to {\it fit} $b(\Phi),c(\Phi)$ and extract
from the data all the information on $\alpha$ and penguin
contributions \cite{quinn}. This requires however a large statistics,
and as in the $\pi^+\pi^-$ case, the measurement of the color
suppressed contribution (here $\rho^0\pi^0$) is mandatory. Based on
some models for the branching ratios \cite{charles} , 
this could require about 3 years of data taking. 

In the first year(s), the approach to this problem will be a 2 body
approach: phase space is integrated , using relativistic Breit-Wigner,
and taking into account interferences between $\rho^+\pi^-$ and
$\rho^-\pi^-$.
 The effects of the penguins are neglected and will induced a
systematic error. A recent analysis of this channel \cite{rhopi}
obtains, for one year of data taking, an effective asymmetry: $\sin 2
\alpha_{eff}=0.26 \pm 0.15 (stat)$ while the generated value (with
penguins) was 0.43. This can give an idea of the induced penguin
shift.

Notice however that for this mode, as for $\pi^+ \pi^-$ very much
depends on what the different measured branching ratios will be.

\subsubsection{Modes studied in \babar}

So far, I have just described 3 analyzes. Many more channels are in fact
studied and Table \ref{tab:modes} summarizes the different modes which
allow a determination of the angles $\beta$ and $\alpha$ of the
Unitarity Triangle.
\begin{table}[ht]
  \begin{center}
    \begin{tabular}{|c|c|c|c|}
\hline
Angle   &	Mode & 	quark process 	&	penguins \\ \hline
$\beta$ & Charmonium $\KS (\KL?)$	& $b \ra c \bar c s$ & $|P|
\ll |T|$ \\
&  Charmonium $ K^{\ast}$	& & \\
& $D^+D^-,~D^{\ast +}D^{\ast -},D D^{\ast}$ & $b \ra c \bar c d$ & $\frac{|P|}{|T|}
? $ \\
& $ \phi \KS $ & $b \ra s \bar s s$ & $ |P| \gg |T|$ \\
\hline
$\alpha$ & $\pi^+ \pi^- $ & $b \ra u \bar u d$ & $\frac{|P|}{|T|} \le
1 $ \\
& $\pi^+ \pi^- \pi^0 (\rho \pi)$	& &$\frac{|P|}{|T|}?$ \\
& $\pi^+ \pi^- \pi^0 \pi^0 (a_1 \pi, \rho\rho)$	& & $\frac{|P|}{|T|}?$ \\
\hline
    \end{tabular}
  \end{center}
\caption{\label{tab:modes} \sl
    Different modes studied in the Collaboration. The last column
indicates the relative contribution between Tree processes and Penguins}
\end{table}
\section{Implications for the Standard Model}

\subsection{\label{sec:sm}Present knowledge of the Unitarity Triangle}

Assuming the Cabbibo angle is known well enough, the observables that constrain
the other 3 parameters of the CKM matrix (\ie~$A,\rho,\eta$ 
in Wolfenstein parametrization) are:
\begin{itemize}
\item
$\vbc$ which is a direct measurement of $A$. A lot of effort both on 
theoretical side (to understand corrections to HQET) and experimental
side has been invested \cite{neubert}. The present knowledge is
(conservatively)\cite{neubert} $\vbc =0.039\pm0.003$
\item
$\vbuvbc$ is obtained from the exclusive or inclusive study of $b\ra u$
decay. In these heavy to light transitions, the theoretical ground is
much less firm. Even with a limited sample, the model-dependent error
dominates and it is reasonable to assume for it a relative error 
as large as 25\%.
\item
$\dmd$ (the $B_d$ mixing frequency) which is a measurement of
$|V_{tb}V_{td}|$. It is now
well known thanks to the LEP time-dependent measurements \cite{LEPWG}. 
In order to extract a relevant CKM quantity, one needs to know
the theoretical parameter $\fbb$ \cite{buras}. There exists a large spread of
estimates for this value depending on the model used (lattices, QCD
sum rules, quark models...). A reasonable {\it range} for these
estimates is $\fbb \in [160,240]\mev$
\item
$\dms$. LEP provided stringent constraints on the mixing frequency for
the $B_s$ meson \cite{LEPWG}. To extract CKM parameters, one needs 
in principle to know a
parameter analogous to the $B_d$ case: $\fbs$. However combined with
$\dmd$, knowing the SU(3) flavor correction:
$\xi_s^2=\frac{\fbb}{\fbs}$ is enough. This latter is better known
from lattices calculations. Still, recent estimates give: 
$\xi_s^2 \in [1.12,1.48]$,
and not knowing more than that, one must, in order to be conservative,
take the upper limit of 1.48. Notice that LEP provided more than a
limit (a set of ``amplitudes''\cite{LEPWG}) and that this information
can be used optimally, as exposed in \cite{bb340}.
\item
$\epsk$ is the measurement of indirect \CP~violation in the kaon
sector. Here the QCD non-perturbative parameter $B_K$ is quite
unknown. A reasonable range is: $B_K \in [0.6,1.]$
\end{itemize}

Before combining these observables, let notice that there is a clear
part of subjectivity for the theoretical parameters used (depending
generally on personal preferences). It is certainly a delicate
matter to estimate which model is right, and what the ``error'' quoted
means? An old Bayesian ghost also appears: not knowing which model is
right is not the same than taking a flat distribution between all estimates.

\babar~has adopted the following way of combining , which is statistically
meaningful \cite{bb340}:

The errors on a quantity are divided into 2 types: experimental errors
are considered to be gaussianly distributed and enter a $\chi^2(A,\rho,\eta)$
estimate. Theoretical parameters are {\it scanned} within reasonable
\footnote {by ``reasonable'' we mean a conservative range obtained after
discussion with many theorists}
ranges: for each scanned parameters, the $\chi^2$ is minimized leading
to an estimate of $A,\rho,\eta$ and for instance a 95\%CL contour in the
$\rho-\eta$ plane. A $\chi^2$ probability cut is applied in order to
check the compatibility between the various observables. If the contour
survives, one then go to the next scanned theoretical parameters, etc.
Knowing the exact theoretical parameter value, one could fix which
contour is the right one. Not knowing it, one takes as a conservative
choice the {\it set} of all the contours as the overall 95\%CL
knowledge of $\rho$ and $\eta$.
Using this method, the present (1998) knowledge of $\rho-\eta$ is
depicted on Fig. \ref{fig:t98} (together with the values used).
\FigPS{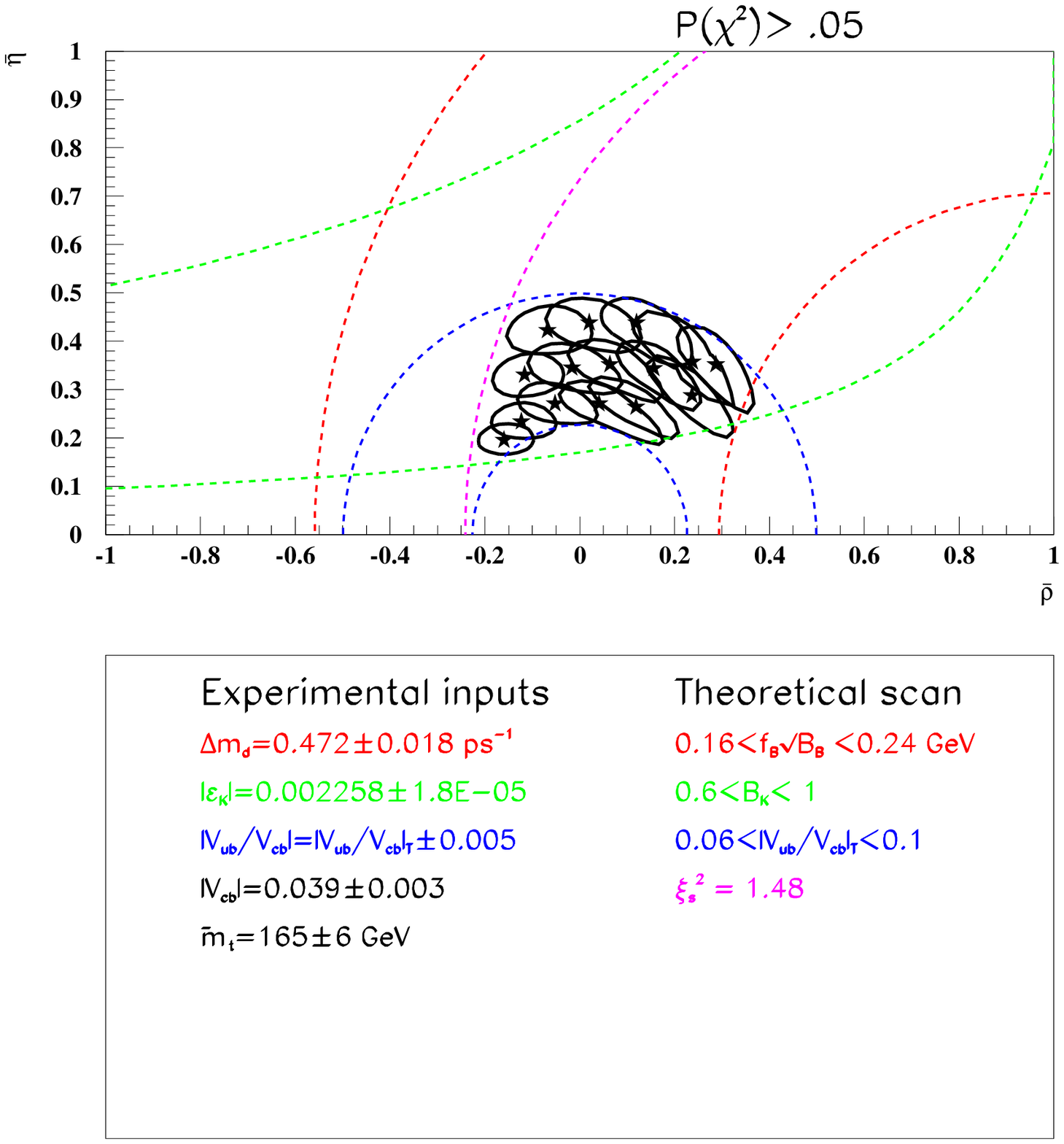}{The 1998 knowledge of the $\rho-\eta$: one circle
represents the 95\% CL obtained when theoretical parameters are
fixed. The set of all contours represents our overall 95\% knowledge
of $\rho\eta$ when scanning the possible range of theoretical
parameters. Also drawn in dashed lines are the constraints provides by
each individual measurement}{6.5in}

\bigskip
Working in another basis ($A,\ssa,\ssb$) , the $\chi^2$ minimization
can be performed and one obtains in the same way the overall 95\% CL
in the $\ssa,\ssb$ plane (Fig. \ref{fig:ss98}).

\FigPS{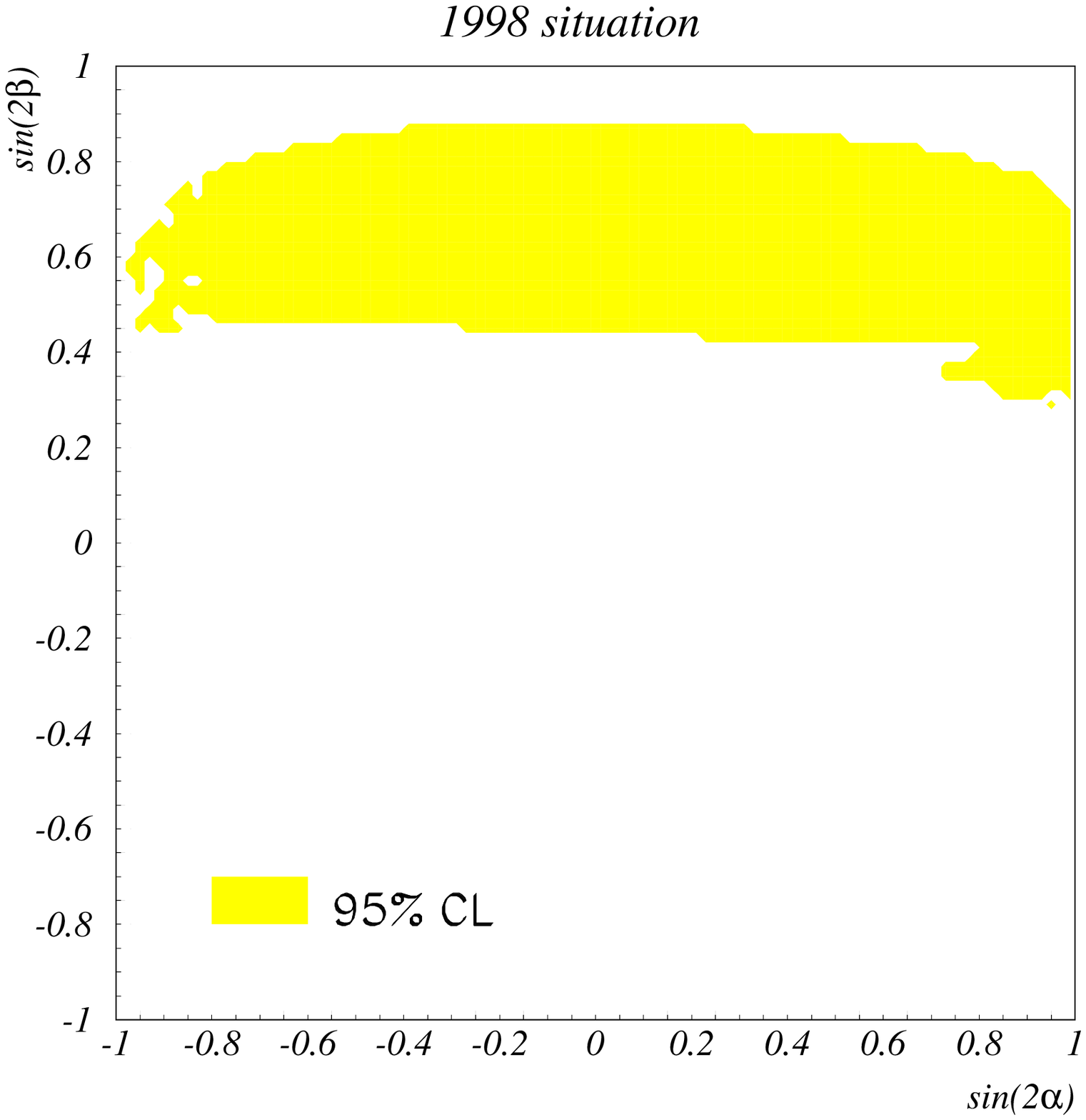}{The 1998 knowledge on $\ssa-\ssb$
 using the same set of values as in Fig \ref{fig:t98} }{5in}.

From this figure, two important remarks arise:
\begin{itemize}
\item
$\ssb$  is presently constrained to lie somewhere on [0.4,0.8](95\% CL) \footnote{Recall
however than this is {\it not a  measurement} and relies on a set of
preferred theoretical parameters: no density distribution can be derived
for it (since the distribution of theoretical parameters is unknown), 
just a range.}. 
The task of modes measuring $\ssb$ will therefore to test the
Standard Model by checking the compatibility with this range.
\item
$\ssa$ is presently unknown. The goal of a B factory is therefore to
{\it measure} this angle.
\end{itemize}

\subsection{What a B factory can bring}

It is presently quite delicate to foresee what the impact of the
B-factory will be in constraining the CKM matrix, since many modes
depend on what the branching ratios, penguins, etc.. actually are.
\babar~has updated \cite{physbook} many of its analyzes with a 
realistic simulation and the full reconstruction program. A 
possible scenario for $30 fb^{-1}$ integrated
luminosity (one year) giving estimated values of these angles is
\beqa
\ssa& = &0.1 \pm 0.1_{exp} \pm 0.2_{th} \\
\ssb& = &0.6 \pm 0.08_{exp}
\eeqa

These numbers represent a reasonable order of magnitude but 
the central values are
completely hypothetical (within the Standard Model). 
Fig \ref{fig:elephant} shows the
impact of such a measurement in the $\rho-\eta$ plane.
\FigPS{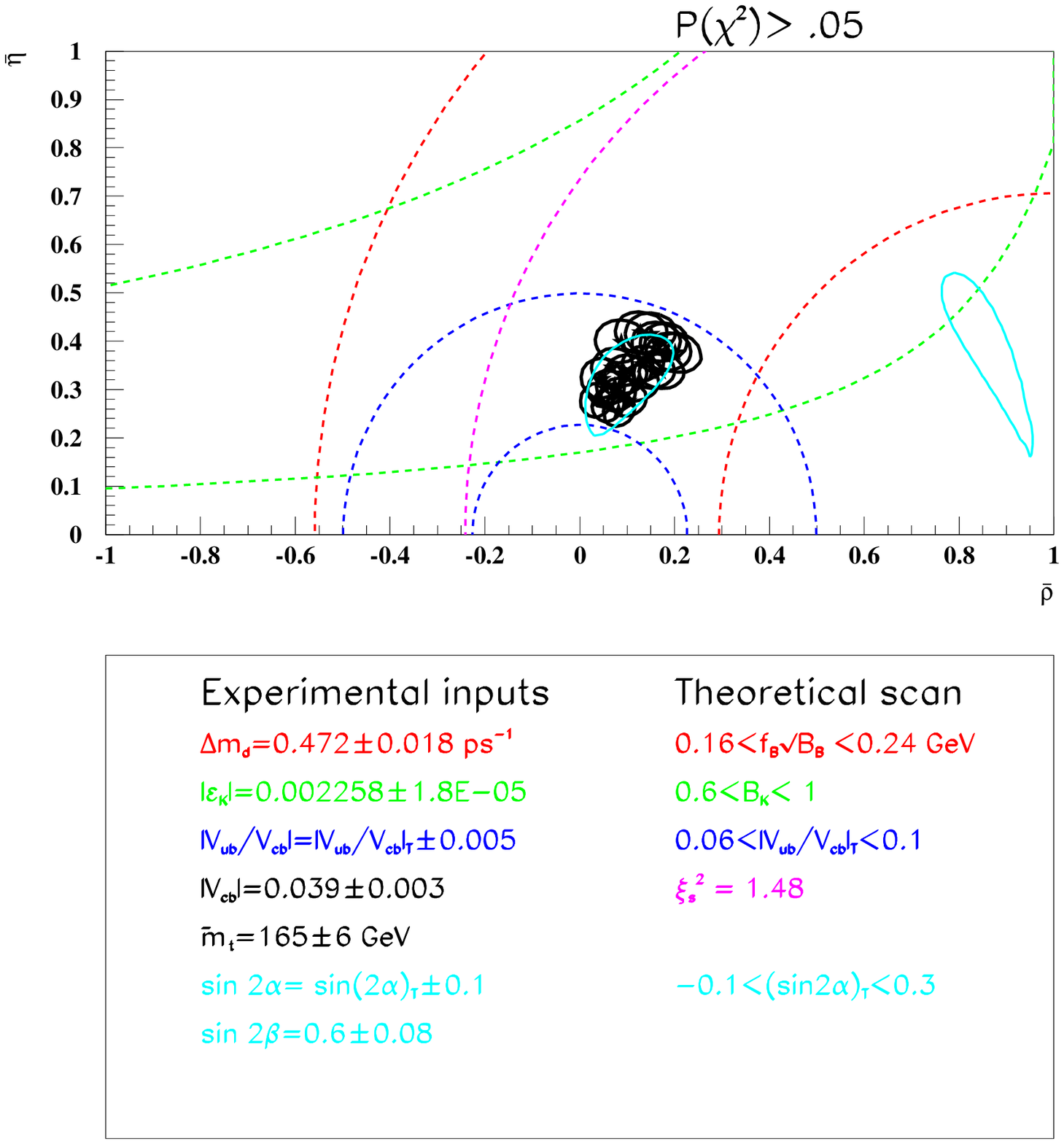}{A possible scenario for the determination of
$\rho-\eta$ after one year of \babar~data taking. Also shown as a new
contour, in light grey (or blue, in color) is the \babar-alone 95\% CL.}{6.5in}

Two final remarks:
\begin{itemize}
\item 
For this combination one neglects the improvement that will appear
with time on the present observables. In particular,
$\vbc,~\vbuvbc,\dmd$ measurements should improve with B-factories.
\item
Just as knowing the sides of the UT allows to already constrain one of
the angle (Fig. \ref{fig:ss98}), in the other way, measuring $\ssa$
and $\ssb$ will allow to {\it measure} in a model independent way, the
theoretical parameters ($\fbb$, $B_K$,$\vbuvbc_{th}$). This happens on
Fig. \ref{fig:elephant} where the combined contour is not better than the 
simple \babar~constraint.
\end{itemize}

\section{Conclusions}

Several tools have been developed in the \babar~collaboration these
last years. In particular:
\begin{itemize}
\item The tagging is now a probabilistic answer of a multivariate
analysis based on discriminating variables. This incorporates optimally
the full information extracted from an event.
\item The \KIN~variable is a golden one for the extraction of an
asymmetry from the data, modelizing all detector effects. It allows to
 present results in a clear way, to optimize selections and combine
the results of many channels (Collaborations?).
\item
A method has been developed to combine observables constraining CKM
parameters in a statistical meaningful way.
\end{itemize}

A B-factory will provide precision tests of the Standard Model.
First of all, it should prove that \CP~violation exists in the
\BzBzb sector. Then many channels will be combined; this require a
deep understanding of the detector. Then, in conjunction with
theoretical work, the relevant CKM angles will be extracted.

\bigskip
Presently $\ssb$ is bound to lie in $[0.4,0.8$](@ 95\%CL) from indirect
measurements (this however implies the choice of theoretical
model-dependent parameters). Therefore the goal of a B factory for this
angle is to check whether the time-dependent asymmetry is compatible
with this range. The golden modes are Charmonium \KS~and
Charmonium $K^{\ast}$.
Since these modes are theoretically well under control, any significant
deviation from this range would indicate a problem (on errors?
theoretical parameters? New Physics?). In particular, if no
asymmetry is observed in $\jpsi\KS$ mode, this would rule out the CKM
mechanism of mixing between 3 fermion families and indicate New Physics.

\bigskip
Presently $\ssa$ is unknown and its determination is 
a challenging task for B-factories.
Very much depends on the values of the branching ratios. In particular if
$\bz\ra\pi^0\pi^0$ (resp. $\bz\ra\rho^0\pi^0$) is measured it allows a
model-independent determination of $\ssa$ from $\pi^+\pi^-$ 
(resp. $\pi^+\pi^-\pi^0$ ). The many accessible rare modes which will
be studied at B-factories will
allow to test different models (as factorization or SU(3) symmetry from
$B(B^0\ra K \pi)$...) and get an insight into the unknown world of penguins.

\subsection*{Acknowledgments}
I am very grateful to R\'emi Lafaye, Sophie Versill\'e, Fran\c{c}ois 
le Diberder and
Carlo Dellapiccola who provided me their supports and work for different
modes. I enjoyed very clear theoretical discussions from J\'erome Charles, 
Olivier P\`ene and Luis Oliver. Finally many thanks to
Marie-H\'el\`ene Schune and Yosef Nir
for discussions, comments and corrections on this work.


\begin{thebibliography}{99}

\bibitem{nirNP}
Y. Nir, talk given at the 18th international symposium on
lepton -- photon interactions, July 28 - August 1, 1997 (Hamburg).\\
hep-ph/9709301

\bibitem{tdr}
\babar~Technical Design Report,SLAC-R-95-457, March 1995.

\bibitem{kin}
S. Versill\'e and F. le Diberder , \babarnote 406\\
S. Versill\'e and F. le Diberder , \babarnote 421
\bibitem{remi}
R.Lafaye, private communication. Work for \cite{physbook}

\bibitem{BKPI}
see Andreas Wolf presentation in this conference.

\bibitem{gronau}
M. Gronau and D. London,\prl65 (1990) 3381. 

\bibitem{grossman}
Y. Grossman and H.R. Quinn, hep-ph/9712306

\bibitem{charles}
J. Charles, private communication.

\bibitem{quinn}
H.R Quinn and A.E. Snyder, \pr D48 (1993) 2139.

\bibitem{rhopi}
S.Versill\'e and F. le Diberder, work for \cite{physbook}.

\bibitem{physbook}
The \babar~Physics Book, SLAC-R-504, in preparation.

\bibitem{neubert}
M. Neubert CERN-TH/98-2, hep-ph/9801269

\bibitem{LEPWG}
``Combined Results on $B^0$ Oscillations:
Update for the Summer 1997 Conferences'',
LEP B Oscillations Working Group,
ALEPH 97-083, CDF Internal Note 4297, DELPHI 97-135,
L3 Internal Note, OPAL Technical Note TN 502, SLD Physics Note 62.

\bibitem{bb340}
S.Plaszczynski and M.H Schune, \babar Note 340.\\
Y. Grossman, Y. Nir, St\'ephane Plaszczynski and Marie-H\'el\`ene
Schune, \np B 511 (1998) 69-84. \\
See also: \cite{physbook}

\bibitem{buras}
For a recent review see:\\
A Buras and R. Fleischer, hep-ph/9704376

\bibitem{shradja}
C. Sachrajda, talk given at the 18th international symposium on
lepton -- photon interactions, July 28 - August 1, 1997 (Hamburg).

\end{thebibliography}
\end{document}